\documentclass[fleqn,10pt]{wlscirep}
\usepackage{subfig}
\title{Fluorographene with Impurities as a Biomimetic Light-Harvesting Medium}

\author[1]{Vladislav Sláma}
\author[1]{Sayeh Rajabi}
\author[1,*]{Tomáš Mančal}
\affil[1]{Faculty of Mathematics and Physics, Charles University, Ke Karlovu 5, 121 16 Prague 2, Czech Republic}

\affil[*]{mancal@karlov.mff.cuni.cz}

\begin{abstract}
We investigate the prospect of using a two-dimensional material, fluorographene, to mimic light-harvesting function of natural photosynthetic antennae. We show by quantum chemical calculations that isles of graphene in a fluorographene sheet can act as quasi-molecules similar to natural pigments from which structures similar in function to photosynthetic antennae can be built. The graphene isles retain enough identity so that they can be used as building blocks to which intuitive design principles of natural photosynthetic antennae can be applied. We examine excited state properties, stability and interactions of these building blocks. Constraints put on the antenna structure by the two-dimensionality of the material as well as the discrete nature of fluorographene sheet are studied. We construct a hypothetical energetic funnel out of two types of quasi-molecules to show how a limited number of building blocks can be arranged to bridge the energy gap and spatial separation in excitation energy transfer. Energy transfer rates for a wide range of the system-environment interaction strengths are predicted. We conclude that conditions for the near unity quantum efficiency of energy transfer are likely to be fulfilled in fluorographene with controlled arrangement of quasi-molecules.
\end{abstract}

\begin{document}

\flushbottom
\maketitle

\thispagestyle{empty}

\section*{Introduction}
Primary processes of photon absorption and exciton transfer in natural photosynthetic antennae have quantum efficiency of almost one \cite{blankenship2013molecular}. This feature is achieved by combination of rapid excitation transfer across spatial extent of the antenna, fast dissipation of excess energy to cross energy gaps between antenna states, and comparatively long life-time of excited states of the constituting light absorbing molecules (pigments). Evolutionary optimization of organisms for survival under a wide range of conditions targets energy transfer efficiency only indirectly. There are other, cheaper strategies to outcompete competitors than optimizing yield of photosynthesis. Accordingly, the overall efficiency in the light-to-useful-biomass energy conversion by plants and algae is only a few percent \cite{blankenship2011comparing}.  However, once the true aim is to harvest as much light energy as possible, the design principles of natural photosynthesis can be applied selectively in hybrid or artificial materials. Artificial bio-inspired design thus starts at the point where natural light-harvesting ceases to be efficient for various biological reasons. Recent advances in understanding photosynthetic energy transport motivate designing biomimetic light harvesting antennae along the quest for alternative resources of energy \cite{calver2016biomimetic, stieger2016biohybrid, gust2012realizing, nozik2002quantum, mcdonald2005solution, semonin2011peak}. The ideal light-harvester is not only highly efficient, but in addition has to enjoy important features such as photo-stability, robustness in mechanical and electronic structure, customizability/adjustability in parameters, as well as responsiveness to a broad range of wavelengths in the solar spectrum \cite{mohseni}. If needed specifically, it also has to be sensitive to particular wavelengths and low intensity of light. Systems inspired by natural photosynthetic machinery provide a road for achieving efficient harvesting of solar energy which is alternative to the more standard photovoltaic concepts. One of the key differences between more traditional photovoltaics and natural photosynthetic antennae is the nature of the excitations transported in these systems. While excitons in the form of bound charges (electrons and holes) move in traditional solid state and hybrid materials \cite{Wenham1996,ORegan1991,Hagfeldt1995}, in photosynthetic antennae, only the excitation energy in form of the so-called Frenkel excitons moves in the system. In photosynthesizing bacteria, for instance, charge separation and electron transfer are restricted to the specifically designed reaction center, and no charges are moved except across a thin cellular membrane \cite{blankenship2013molecular,ruban2012photosynthetic}.     

The attempts towards building artificial light harvesting systems have mainly been centered around integrating natural antennae into inorganic devices \cite{calver2016biomimetic,stieger2016biohybrid}, synthesizing macromolecules inspired by the structure of photosynthetic antennae \cite{gust2012realizing}, as well as photovoltaic devices based on nano-systems such as quantum dots \cite{nozik2002quantum, mcdonald2005solution, semonin2011peak}. The present work tries to revive the hope for constructing biomimetic light harvesting antennae in a bottom-up process, using design principles learned from nature, starting from the molecules as the building blocks, up to desired large-scale structures. We are inspired by the established explanation of the photosynthetic energy transfer efficiency which stresses interplay between exciton delocalization, resonance between electronic energy gaps and environmental spectral density, and funnel-type energy landscape of the antenna \cite{may2008charge,van2000photosynthetic,valkunas2013molecular}. 

\begin{figure} 
  \centering
 \subfloat[]{\includegraphics[width=0.16\textwidth]{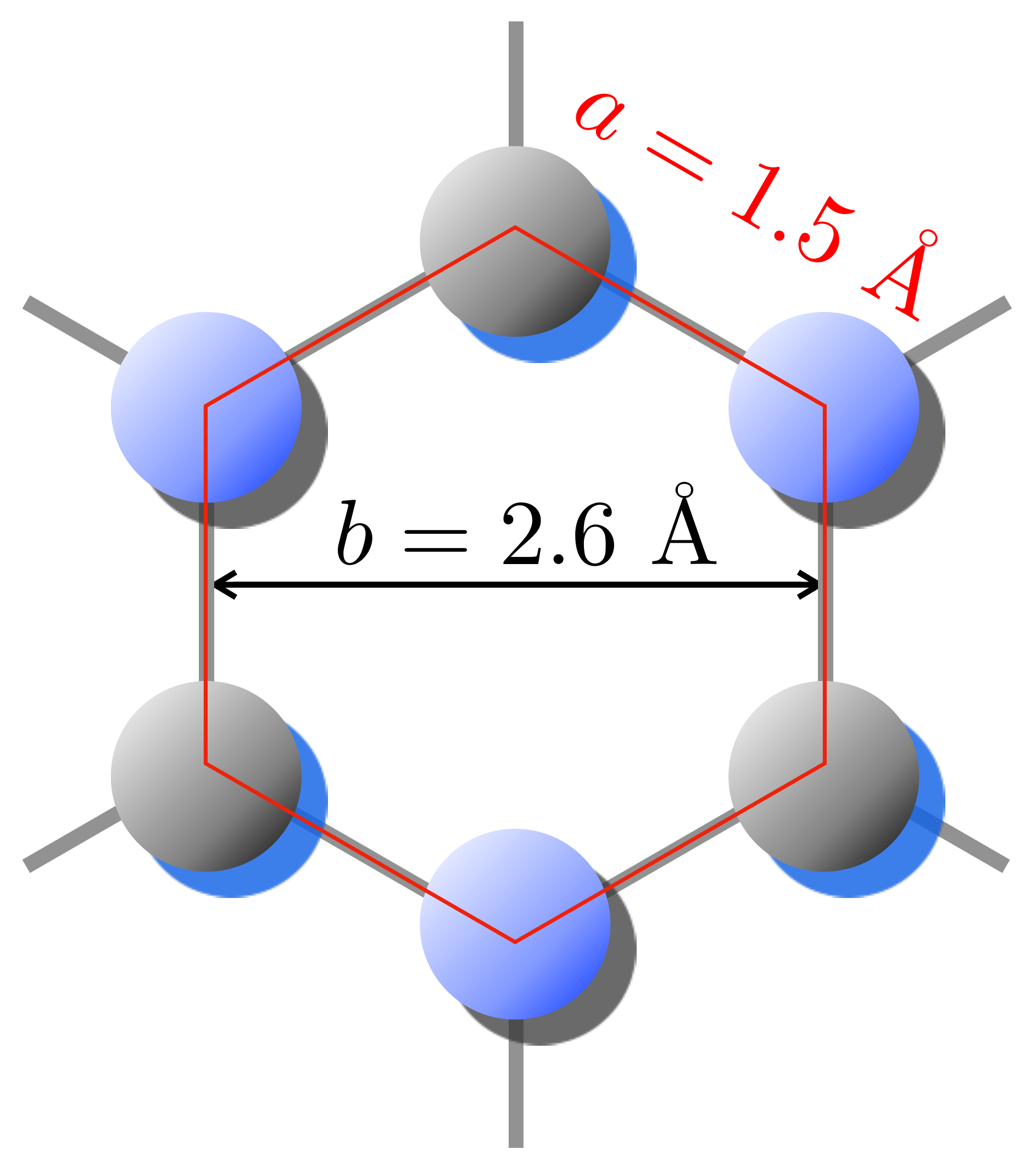}\label{fig:FG-Lattice-hex}}
  \subfloat[]{\includegraphics[width=0.27\textwidth]{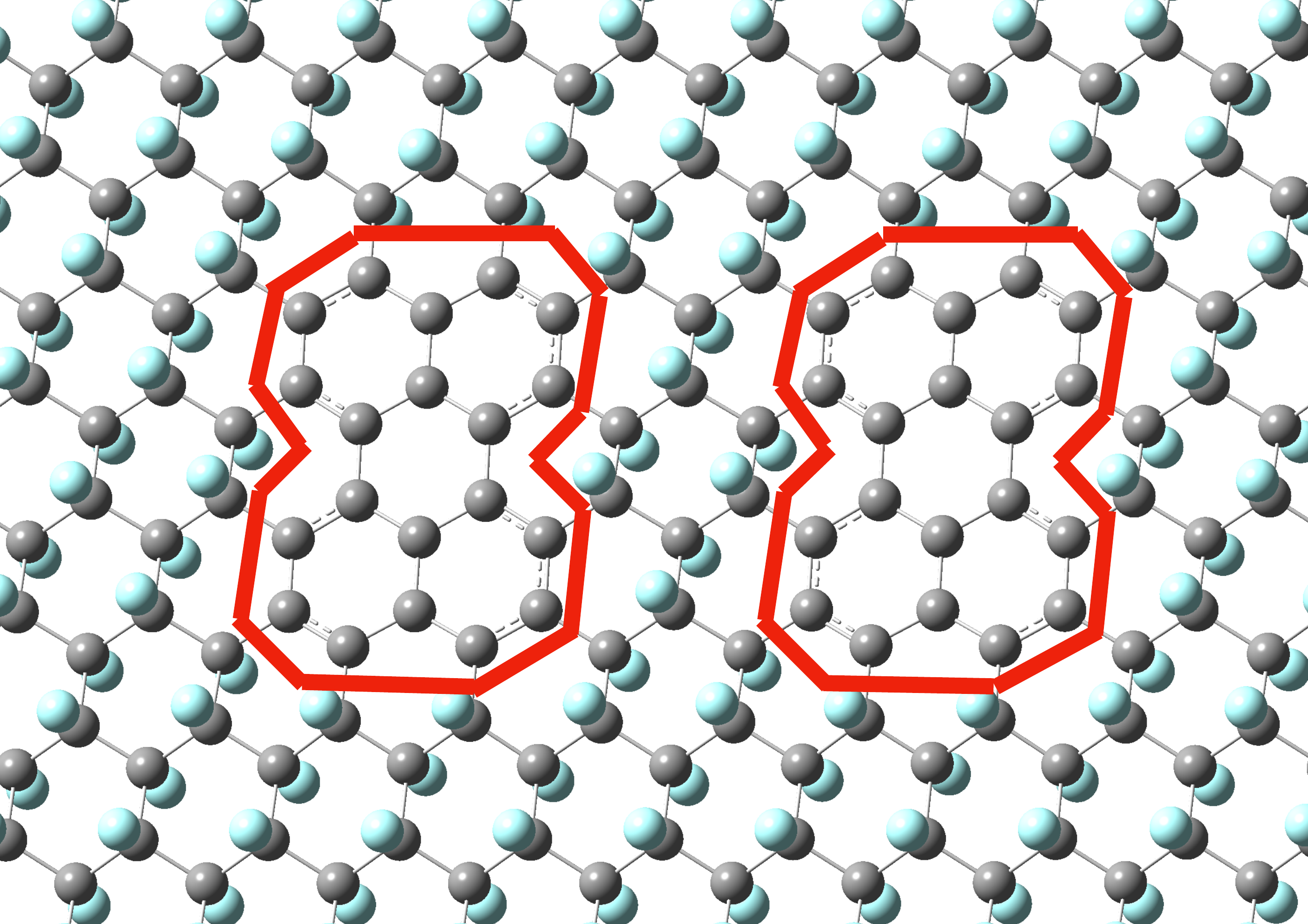}\label{fig:FG-Per-defects}}
  \subfloat[]{\includegraphics[width=0.24\textwidth]{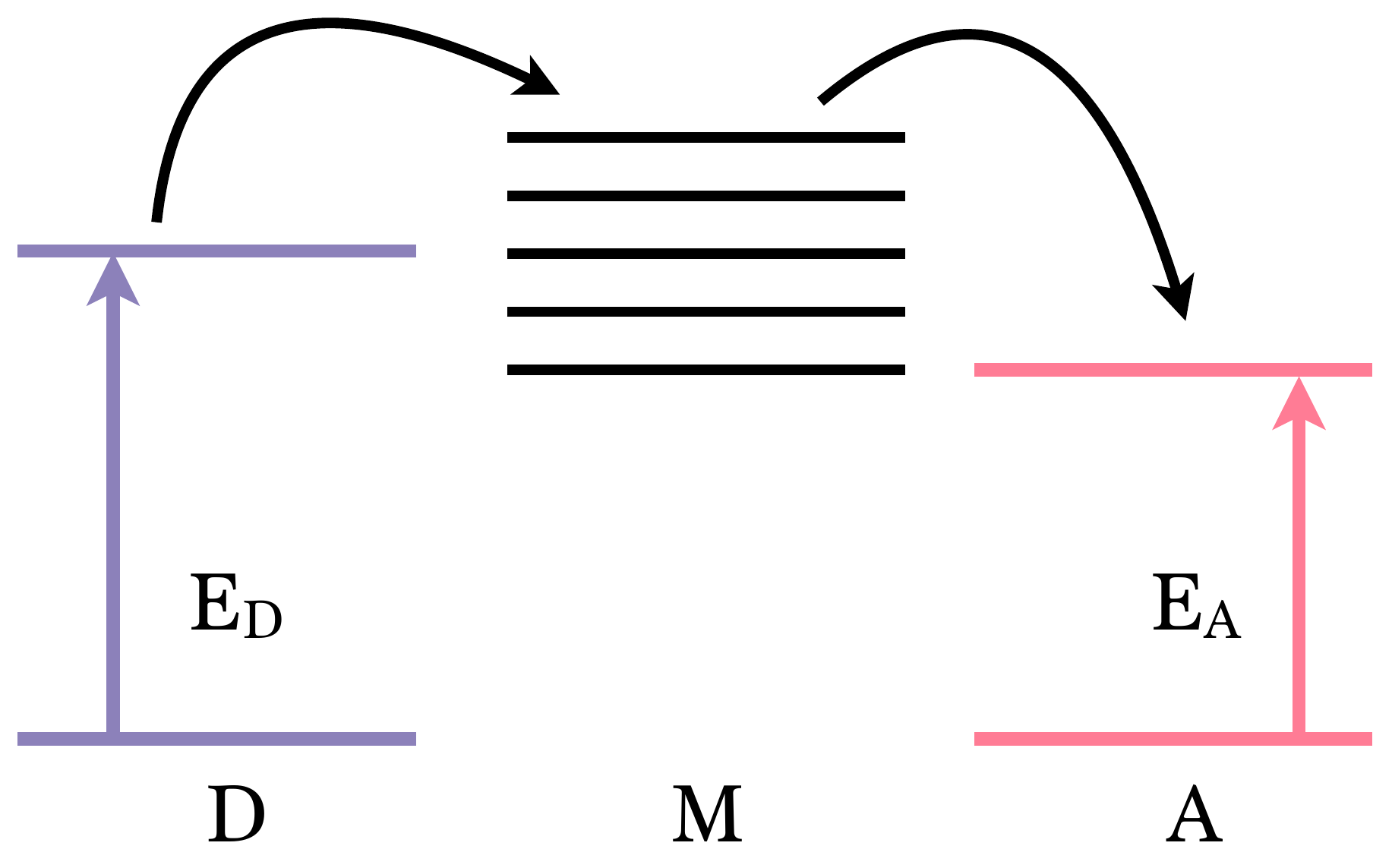}\label{fig:ET-diagram}}
  \subfloat[]{\includegraphics[width=0.29\textwidth]{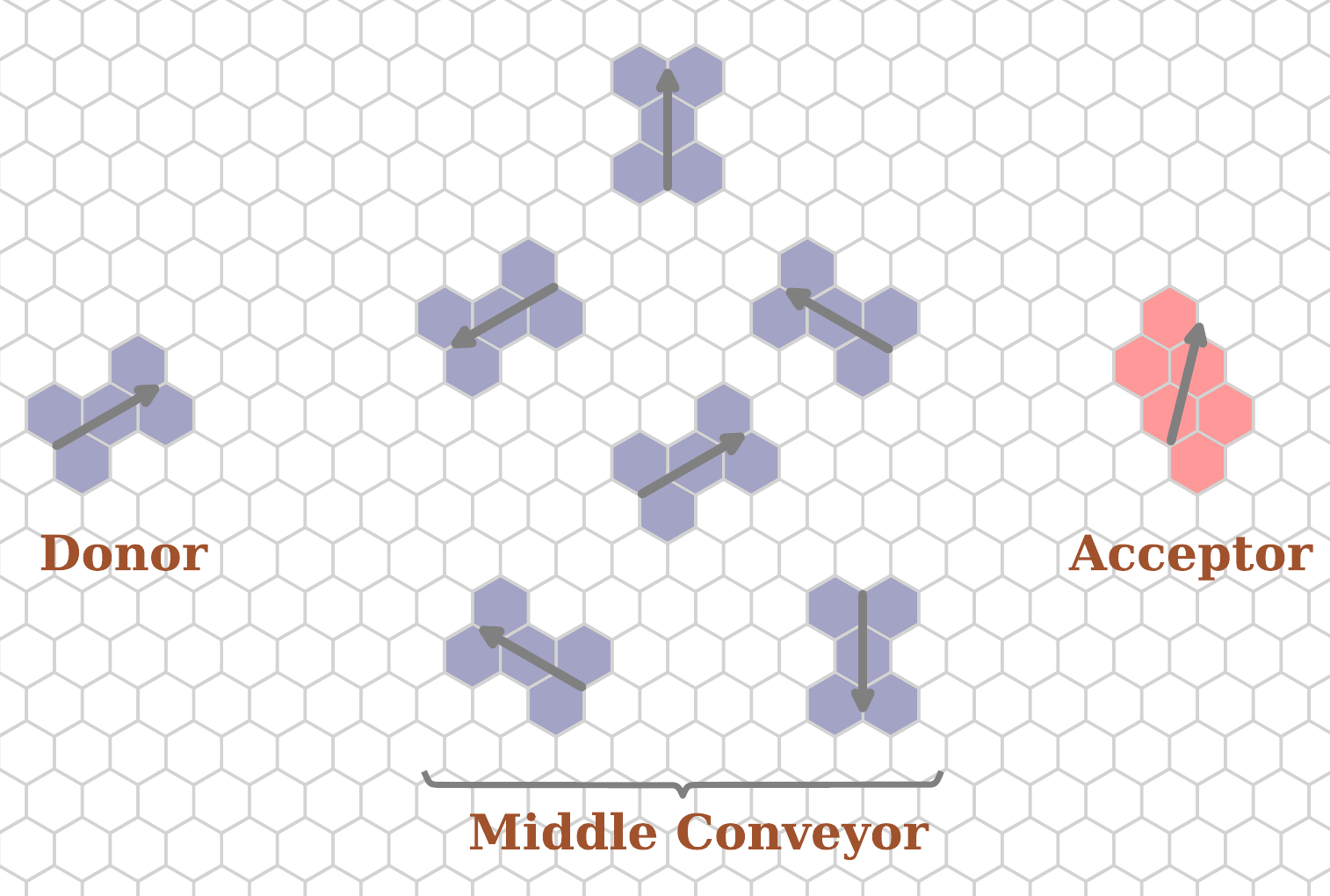}\label{fig:DMA1}}
\caption{(a) Hexagonal cell of fluorographene lattice. (b) Two perylene-like defects on the fluorographene lattice. (c) The idea of energy funneling in the antenna: excitation energy transfer from donor to acceptor through a middle aggregate which provides quasi-resonant states to enhance the transfer. (d) Example of an 8-site antenna on fluorographene lattice:  donor and middle conveyor are made up from perylene-like quasi-molecules, and anthanthrene-like quasi-molecules constitute the acceptor. The arrows show the direction of the quasi-molecule transition dipole moments.}
\end{figure}

Our material of interest to serve as the medium for light-harvesting is fluorinated graphene\cite{sofo2007graphane}, fluorographene $(\text{CF})_{\text n}$ -- 1:1 ratio of fluorine to carbon (hexagonal cell, Figure \ref{fig:FG-Lattice-hex}) -- whose electrical, optical and chemical properties have been subject to many recent studies \cite{robinson2010properties, nair2010fluorographene,zbovril2010graphene,karlicky2013band,paupitz2012graphene}. With their exceptional electronic and mechanical properties, graphene and its derivatives have highlighted new prospects for theoretical research and technological advances. Fluorographene (FG) is a thermally stable insulator with large energy gap above 3 eV \cite{jeon2011fluorographene, nair2010fluorographene} 
which makes it transparent to almost entire solar spectrum \cite{nair2010fluorographene}. Light harvesting in the visible range and corresponding excitation energy transfer are not directly possible by means of pure FG due to its wide energy gap. However, when some adjacent fluorine atoms are removed from the material, isles of graphene with different shapes could form on the FG lattice, e.g. perylene-like defects in Figure \ref{fig:FG-Per-defects}. These defects or impurities in pure FG have $\pi-$conjugated electron orbitals, and by chemical intuition they should exhibit electronic transitions in the FG gap. In this work, we report on the employment of such FG impurities towards the end of light harvesting and energy transport. Graphene-like defects play here the role of chromophores in plant or bacterial photosynthesis. From a solid state physics perspective, graphene isles provide impurity states within the energy gap of FG. These shorten the energy gap to allow visible light harvesting and aid fast excitation energy transport (Figure \ref{fig:ET-diagram}).

In studying two-dimensional (2D) materials, the chemical and solid state physics perspectives inevitably meet. There are known solid state techniques to efficiently study symmetric 2D materials \cite{FoaTorresBook2014, WolfBook2014, KatsnelsonBook2012}. However, systems from which we borrow design principles, i.e. the natural light harvesting antennae, usually lack translational symmetry, as they are composed of small, rather independent complexes. These complexes have dimensions of only few nanometers, and most of their important function is determined by their local properties on the nanometers scale. Precisely controlled manipulation of impurities in FG sheets is currently not readily available. To realize experiments, one would have to rely on defects occurring naturally in nearly pure FG. Also here, partial defluorination is unlikely to show symmetric distribution of defects. For these reasons, it is reasonable to approach studying the prospects of utilizing impurities in FG for light-harvesting means from a molecular, i.e., chemical perspective. The molecular approach to calculation of the optical and transport properties of this material can directly benefit from the experience drawn from the field of study of natural photosynthetic systems. 

The workhorse of the theory of natural photosynthetic antennae is the so-called Frenkel exciton model \cite{van2000photosynthetic}. Frenkel exciton model is the simplest level of application of configuration interaction method of quantum chemistry on a molecular system composed of molecules with zero mutual differential overlap of orbitals \cite{seibt2016optical}. The condition of zero differential overlap is not only an approximation which yields the problem of construction of an effective electronic Hamiltonian and wavefunction tractable (for instance, the need for anti-symmetrization of electronic wavefunction can be effectively avoided). It also reflects important physics of the problem, namely the lack of electron exchange between molecules constituting the photosynthetic antennae. Electrons are effectively bound to the molecules, no charge is transferred during the excitation energy transfer, and excitations move only by resonant exchange of excitation energy mediated by Coulombic coupling between the molecules. Given that charge transfer states -- enabled by non-zero differential overlap -- are often implicated in quenching of excitations in photosynthetic self-regulation processes \cite{Wahadoszamen2014, Beddard1976}, zero differential overlap between neighbouring molecules can be counted among the design principles of efficient light-harvesting by photosynthetic pigment-protein antennae. 
Despite a forbidden electron exchange, excited states of Frenkel exciton systems exhibit delocalization due to coupling between transitions on different molecules. Local fluctuations of energy gap, i.e. the electron-phonon- or more generally system-environment coupling, effectively localize excited states in a process termed dynamic localization \cite{Renger2004}. The resonance coupling and system-environment coupling thus compete in a given excitonic system, and they determine the system's effective energy level structure.

We propose to build artificial light-harvesters by suitable organization of defects on FG. We show below that the $\pi$-conjugated electronic states of these defects match very well the $\pi$-electron states of the corresponding isolated hydrocarbon molecules, suggesting that these graphene-like defects on the FG sheet can be treated as quasi-molecules, or \textit{moleculoids}. In the rest of the paper, we will use the term moleculoid with the same meaning as quasi-molecule, graphene-like impurity or defect in FG. We will also use the name of corresponding molecules for the moleculoids. Thus we will refer to a perylene shaped isle of carbon atoms with $\pi$-conjugated bonds in FG by the term perylene moleculoid.  Below, we demonstrate that when the moleculoids are placed next to each other, separated by the minimum one layer of fluorinated carbon atoms, they exhibit negligible differential overlap of the $\pi$-conjugated electronic orbitals, i.e. negligible electron transfer. They generally satisfy the conditions for treatment with Frenkel exciton model \cite{blankenship2013molecular}. Our quantum chemical (QC) calculations verify the stability of these defects and confirm that they have transition energies within the energy gap of FG. We also show how certain arrangements of even a few moleculoids on the FG lattice lead to a more efficient energy transfer between two collections of moleculoids playing the roles of donor and acceptor of excitations. We demonstrate the idea of excitation energy transfer and occurrence of energy funneling in a model light harvesting antenna made up from perylene and anthanthrene moleculoids. Perylene and anthanthrene are the smallest symmetric aromatic hydrocarbons built out of hexagonal (benzen) rings which have the excitation energies smaller than FG energy gap and the lowest electronic excited state allowed for the optical excitation. They represent perhaps the smallest moleculoids to be employed in constructing interesting model light-harvesting antenna. Larger moleculoids may turn out to be more suitable for construction of a practical antenna. However, for computational reasons, smaller moleculoids are more apt for the present proof of the principle. 

\section*{Results}

\subsection*{Electronic States of Impurities in Fluorographene}

\begin{figure}[tbp]
  \centering
  \subfloat[]{ \raisebox{-0.5\height}{\includegraphics[width=0.14\textwidth]{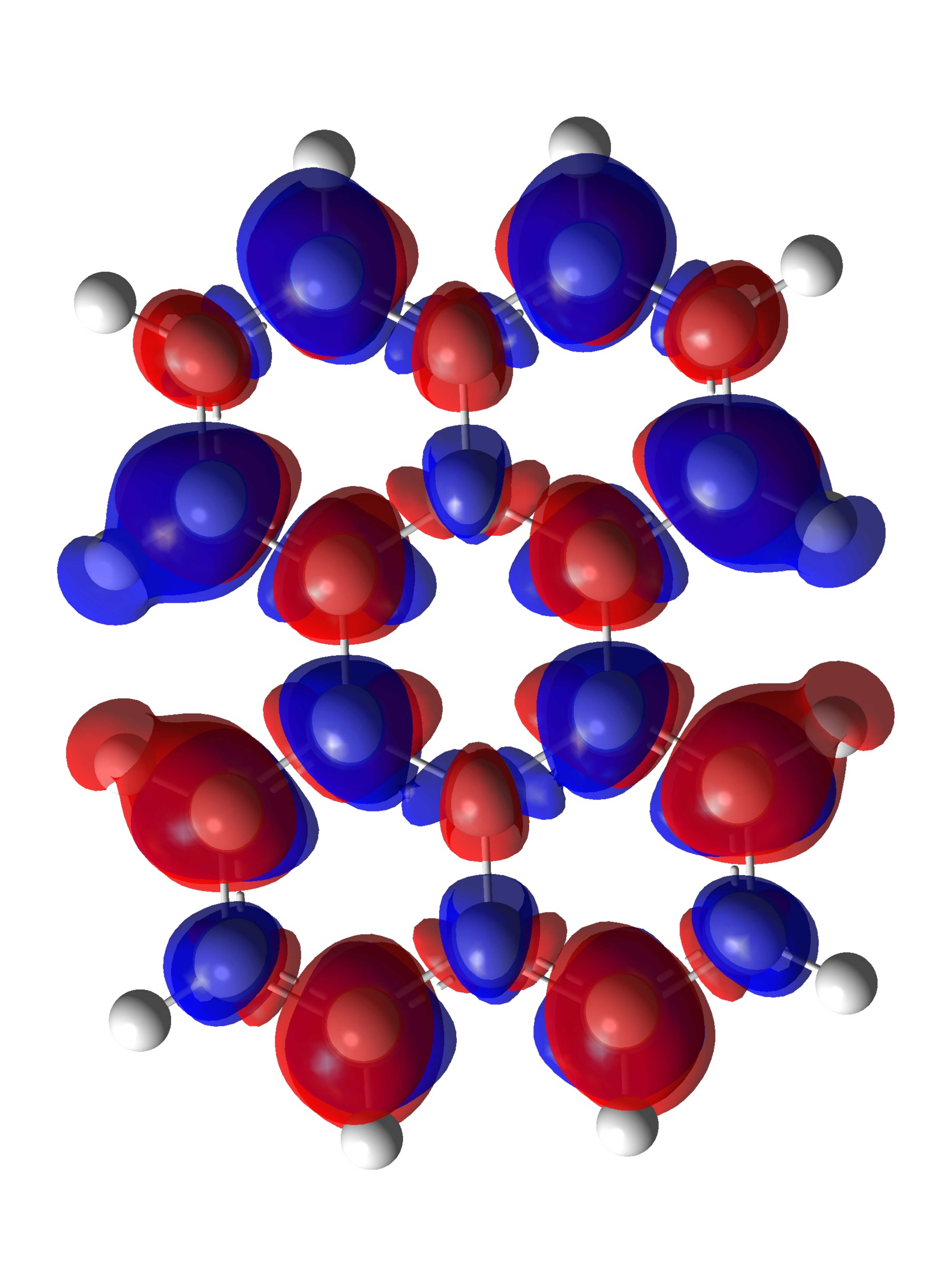}}\label{fig:per_TrDens}} 
  \subfloat[]{ $\vcenter{\hbox{{ \includegraphics[width=0.30\textwidth]{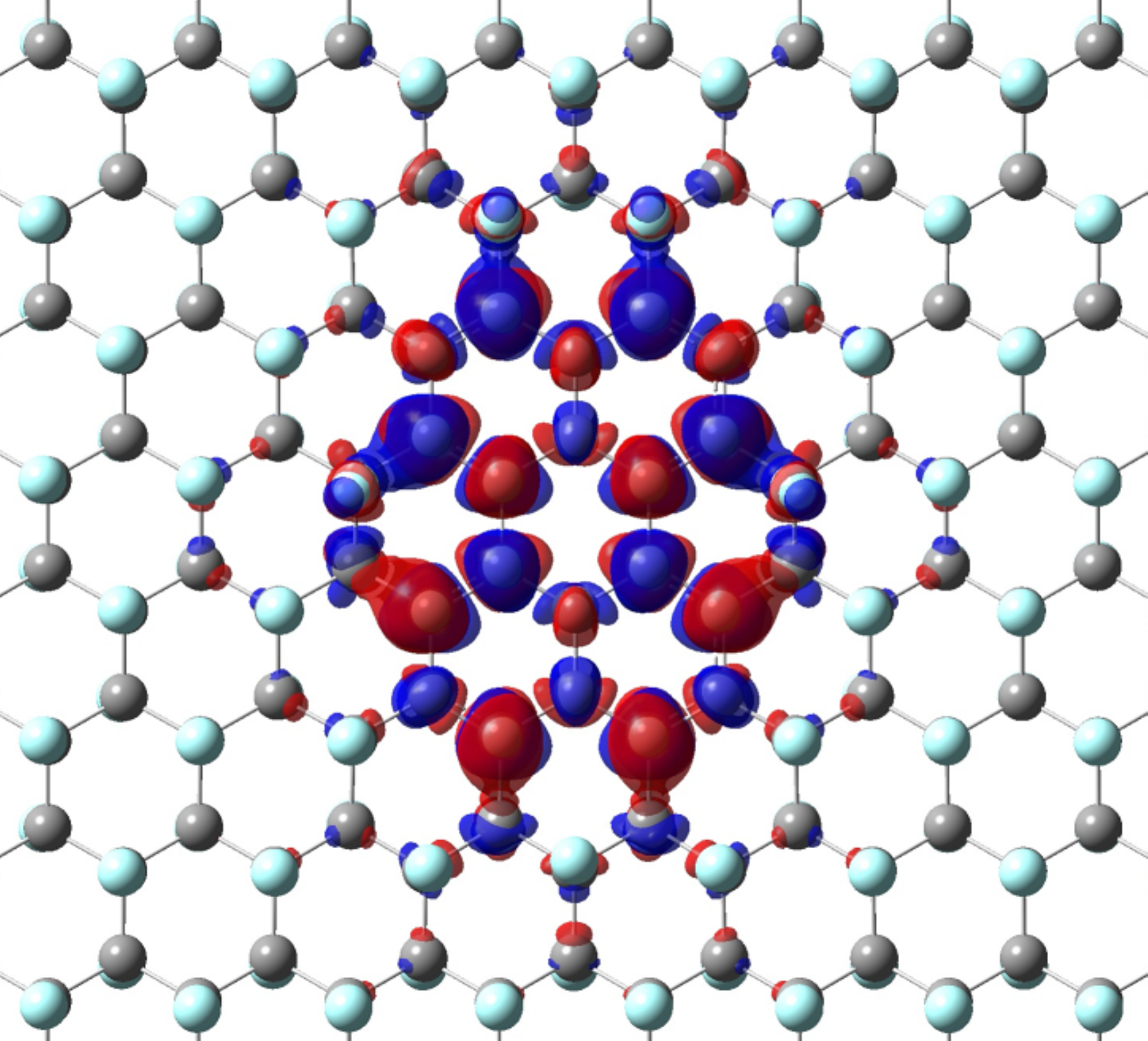}\label{fig:FGper_TrDens}  }}}$ } 
    \subfloat[]{ $\vcenter{\hbox{{ \includegraphics[width=0.19\textwidth]{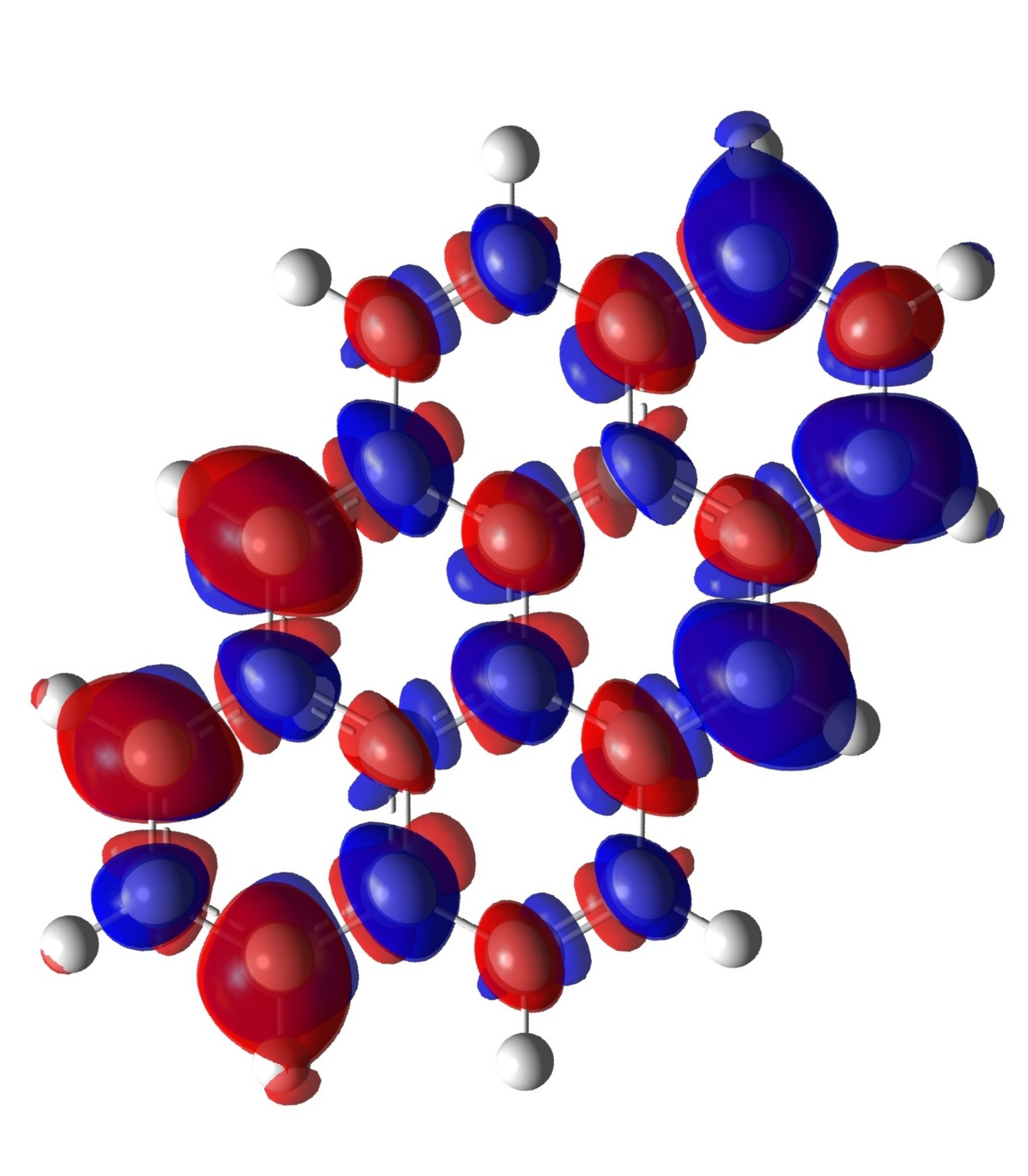}\label{fig:anth_TrDens} }}}$ }
  \subfloat[]{ $\vcenter{\hbox{{ \includegraphics[width=0.26\textwidth]{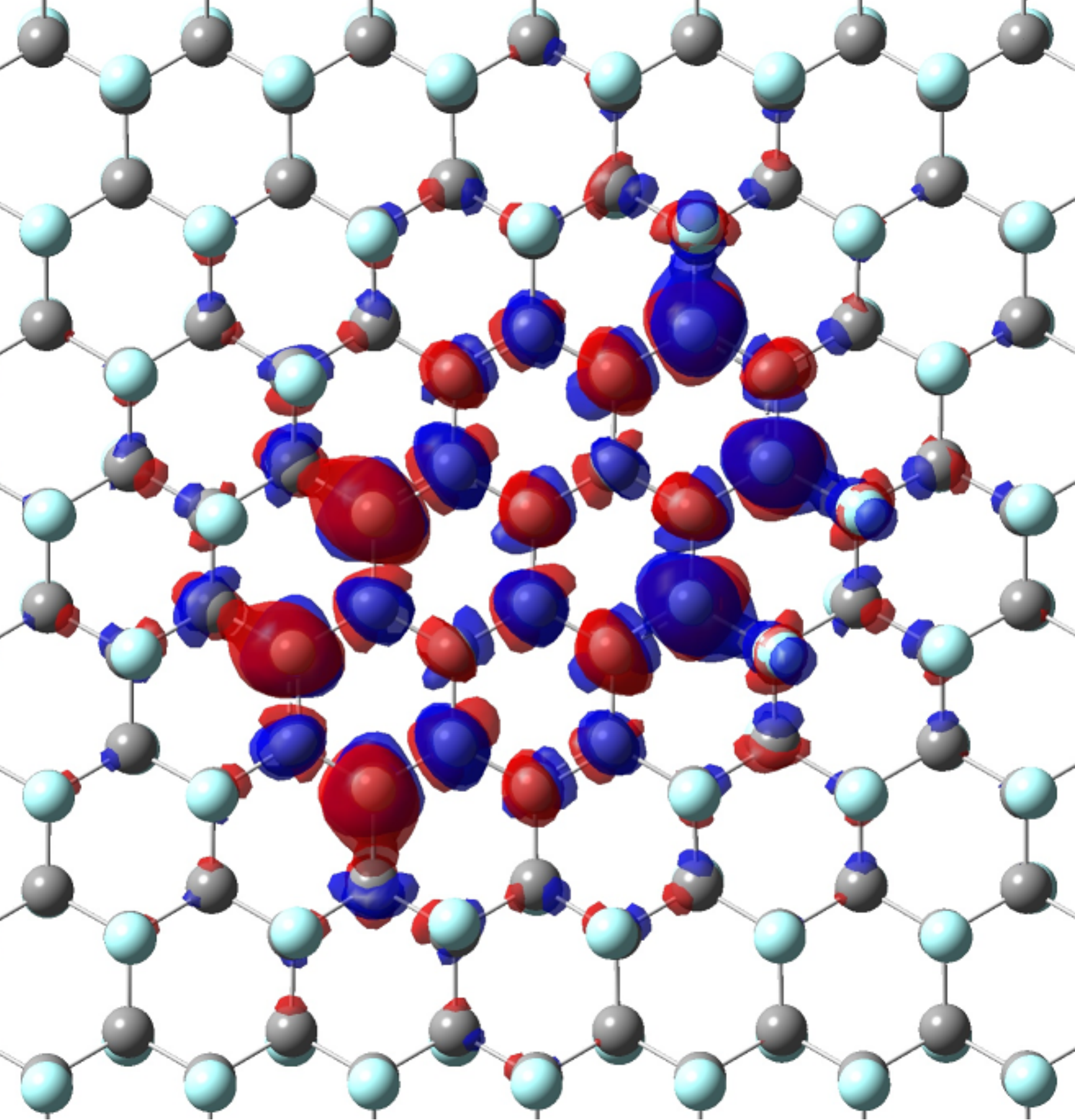}\label{fig:FGanth_TrDens}} }}$}
  \caption{\label{fig:FG_TrDens} Transition density between electronic ground state and the lowest electronically excited states obtained from quantum chemistry calculation for isolated perylene molecule (a), perylene moleculoid on fluorographene (b), anthanthrene molecule (c) and anthanthrene moleculoid on fluorographene (d). Transition density is represented as iso-surface with density value of 0.0008.}
\end{figure}

\begin{figure}[tbp]
  \centering
  \subfloat[]{\includegraphics[width=0.39\textwidth]{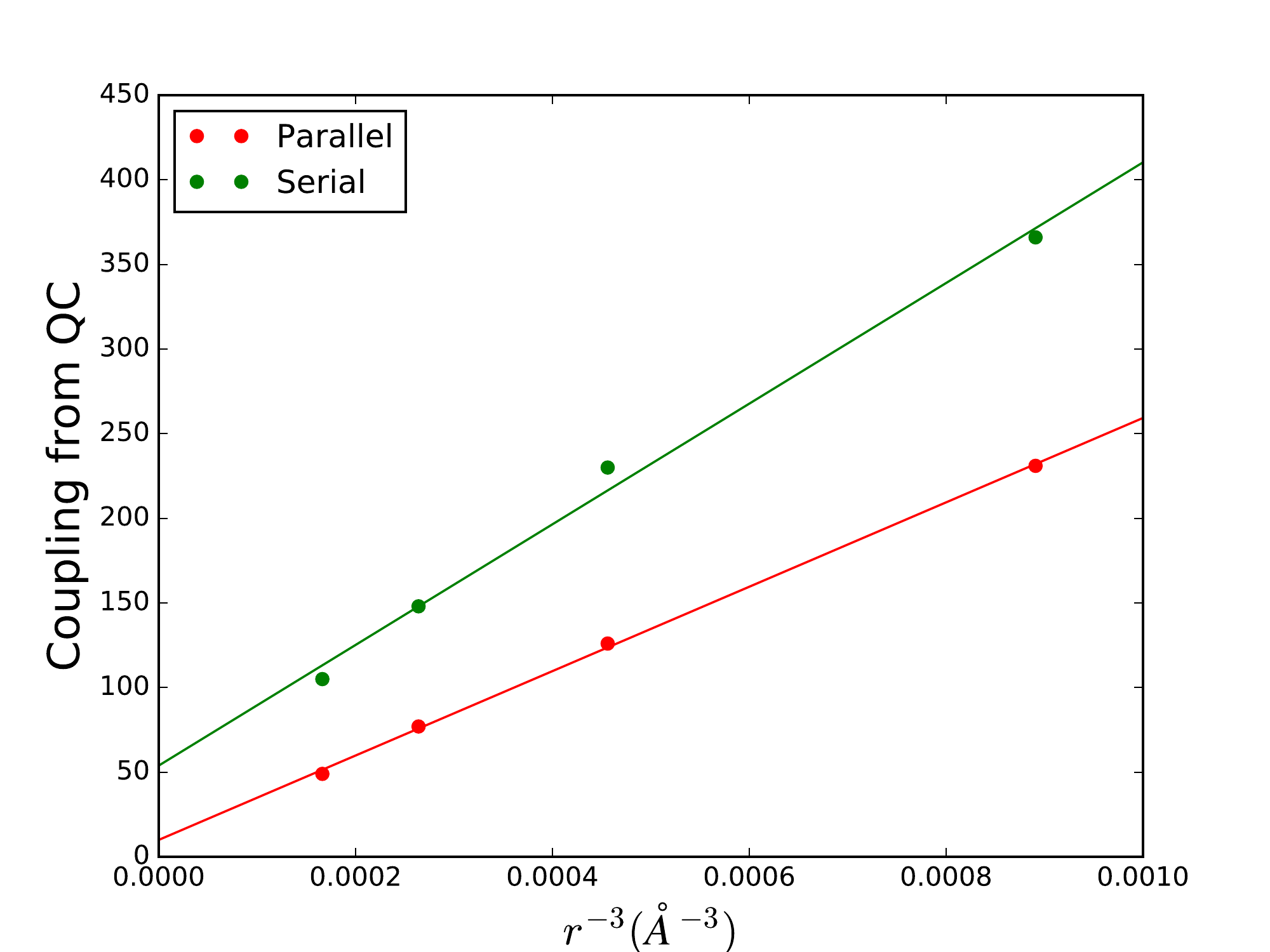}\label{fig:rel_perm_a}}
  \subfloat[]{\includegraphics[width=0.38\textwidth]{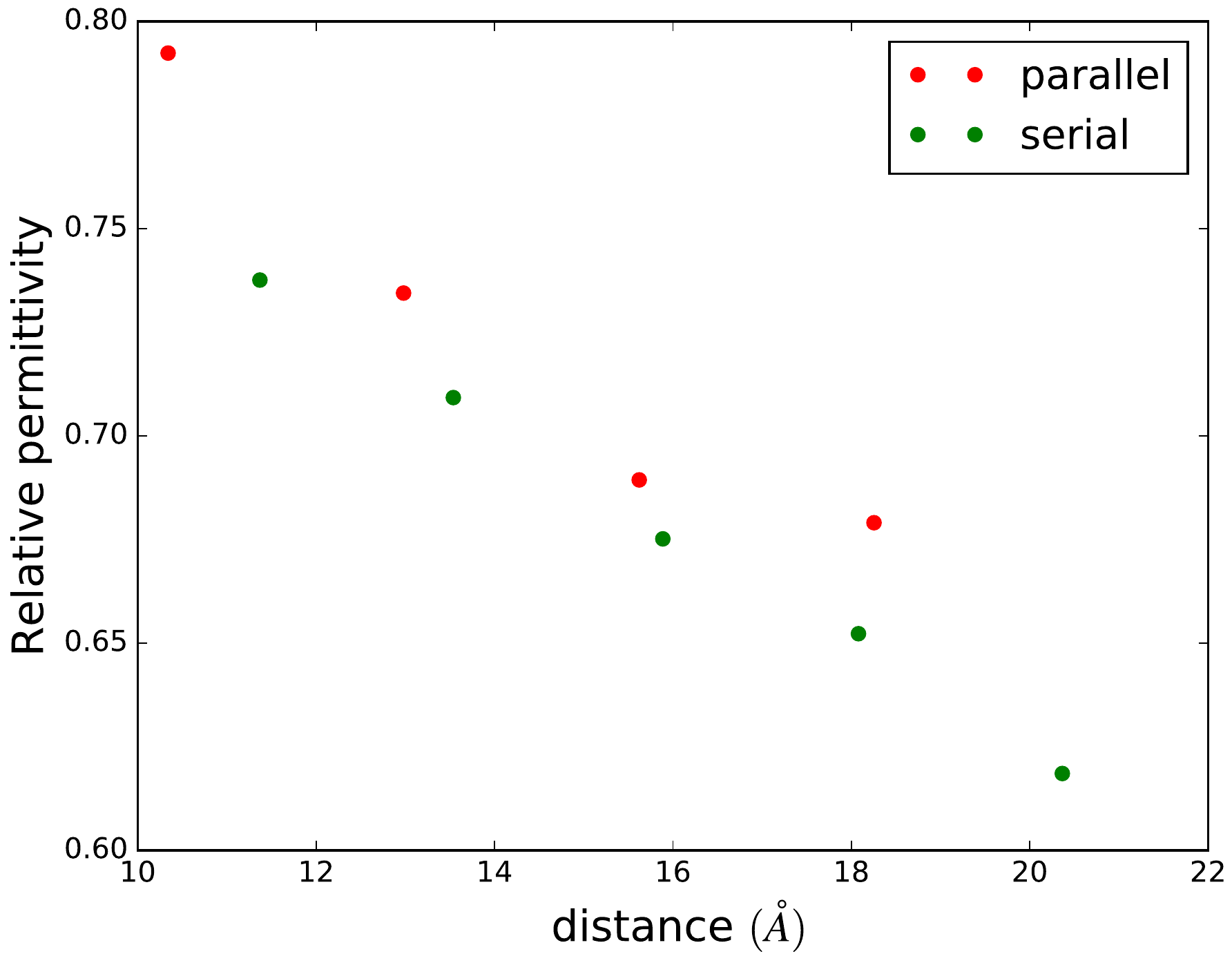}\label{fig:rel_perm_b}}
  \caption{(a) Resonance couplings calculated by quantum chemistry from excited states splitting for two mutual orientations (parallel and serial) of dipole moments with various distances between the centers of perylene moleculoids. Linearity confirms the validity of dipole-dipole approximation. (b) Effective relative permittivity of fluorographene as a function of distance between two perylene moleculoids with both parallel and serial transition dipole moments.}
\end{figure}

FG has a non-planar chair conformation whose 2D image suffices for our analysis. Hence, we will work with a regular lattice with a hexagonal cell as shown in Figure \ref{fig:FG-Lattice-hex}. The average lattice constant is taken $a = 1.5$ \AA. Taking $b$ as the separation of parallel edges of the hexagon, one finds $b = a\sqrt{3}$ with average value of $2.6$ \AA. See Supporting Information (SI) for details. QC calculations show the transition between the electronic ground and the first excited states of a perylene moleculoid to exhibit transition dipole moment along its longer mirror symmetry axis. This is intuitively clear from the symmetry of the moleculoid or its corresponding molecule. For anthanthrene, the relevant transition dipole moment makes a sharp angle with the longest axis of the molecule/moleculoid (Figure \ref{fig:DMA1}). On the FG lattice, transition dipole moment of perylene moleculoid can thus take six possible directions: $\pm\pi/6, \pm\pi/2, \pm5\pi/6$.  Since anthanthrene is chiral, there are twelve possible directions for its transition dipole moment. We take one of the orientations, along $(b,7a)$, for anthanthrene moleculoids to constitute the acceptor in our model (Figure \ref{fig:DMA1}).

To place the moleculoids on the lattice, regions in which orbital overlap between each moleculoid and its neighbours can occur must be excluded. Reason for this spatial separation is to have no overlap between $\pi$-conjugated states of the individual moleculoids. When overlap is present, the whole system has to be treated as a single moleculoid. This {\it forbidden region} depends on the relative orientation of the transition dipole moments. For instance, the minimum distance between the centers of two perylene moleculoids with parallel dipole moments ($\vec{d}$), and if $\vec r_{ij}\perp\vec d_{i , j}$, is $4b = 10.4$ \AA \ (Figure \ref{fig:FG-Per-defects}). 

To construct electronic Hamiltonian of the moleculoids on FG sheet, we employ the Frenkel exciton model. Each moleculoid is considered as a two-level system, with electronic ground state and the first allowed electronic excited state whose properties are determined by QC calculations. Represented in the basis of collective singly excited electronic states of an aggregate of moleculoids, the (electronic) Frenkel Hamiltonian reads as
\begin{align}
H_e = \sum_i E_i |i\rangle\langle i| + \sum_{i,j} J_{ij}(|i\rangle\langle j| + |j\rangle\langle i|),
\end{align}
where $E_i$ is the excitation energy of the $i$th moleculoid (we set its ground state energy to zero), and $J_{ij}$ denotes the Coulombic resonance coupling between transitions on moleculoids positioned at sites $i$ and $j$. The collective states
$|i\rangle$ read
$|i\rangle = |g_1\rangle \dots |e_i\rangle \dots |g_N\rangle$, $i=1, \dots, N$,
where $|g_i\rangle$ and $|e_i\rangle$ are electronic ground state and electronic excited states of the $i$th molecule, respectively. The fact that we consider only single exciton states of the moleculoid aggregate is in line with the situation in natural photosynthetic systems. We assume that the density of moleculoids as well as their transition dipole moments will be similar to the density and transition dipole moments of chromophores in naturally occurring systems. They will also be subject to the same illumination intensity. It is well known that to describe modern time-resolved laser spectroscopy on such systems, one is required to work with up to two-exciton manifold of states, and for the description of transport properties, only single-exciton manifold is needed \cite{mohseni,valkunas2013molecular}. 

To determine the parameters of the Frenkel exciton Hamiltonian, and to verify that properties of the moleculoids are as expected, we employ standard QC methods (see Methods section and SI) to sections of FG sheets with model impurities. 
The defects were created by dissociation of fluorine atoms from even number of neighbouring carbons on FG surface in order to create structures without radical or ionic character. Between two carbons with fluorine vacancy, double bond is formed. These types of structure have lower formation free energy than the same number of single fluorine vacancies scattered around FG surface \cite{dubecky2015reactivity}. Comparison of ground state energies of optimized structures of three types of FG systems with defects revealed that compact defects are more stable and energetically favorable than more disordered defects (see SI). This is in agreement with previous findings\cite{dubecky2015reactivity}, where it is proposed that fluorine atoms tend to dissociate in the vicinity of already formed fluorine vacancy due to the lower formation free energy.

Inspection of molecular orbitals of compact moleculoids revealed confinement of $\pi$-molecular orbitals in the area of the defect, as was also reported earlier\cite{ribas2011patterning}, whereas the sigma molecular orbitals are delocalized through the whole FG sheet. Due to the localization of $\pi$-conjugated states, the lowest optical transitions should be also well localized on the moleculoid. This was confirmed by QC calculation of excited state properties of perylene and anthanthrene moleculoids. Transition density of FG sheet with a moleculoid is well localized in the area of the moleculoid with only small leakage into FG, and it is similar to the transition density of corresponding isolated molecule. Figure \ref{fig:FG_TrDens} presents the transition density for the perylene and anthanthrene molecules and the corresponding moleculoids. Localization of moleculoid orbitals supports the general idea that moleculoids in FG sheet can be used for conversion of visible light into Frenkel excitons and subsequent excitation energy transfer in a way similar to the function of pigments in natural light harvesting complexes.

\begin{figure}[tbp]
  \centering
  \subfloat[Coupling map of two upward perylene-like moleculoids]{\includegraphics[width=0.44\textwidth]{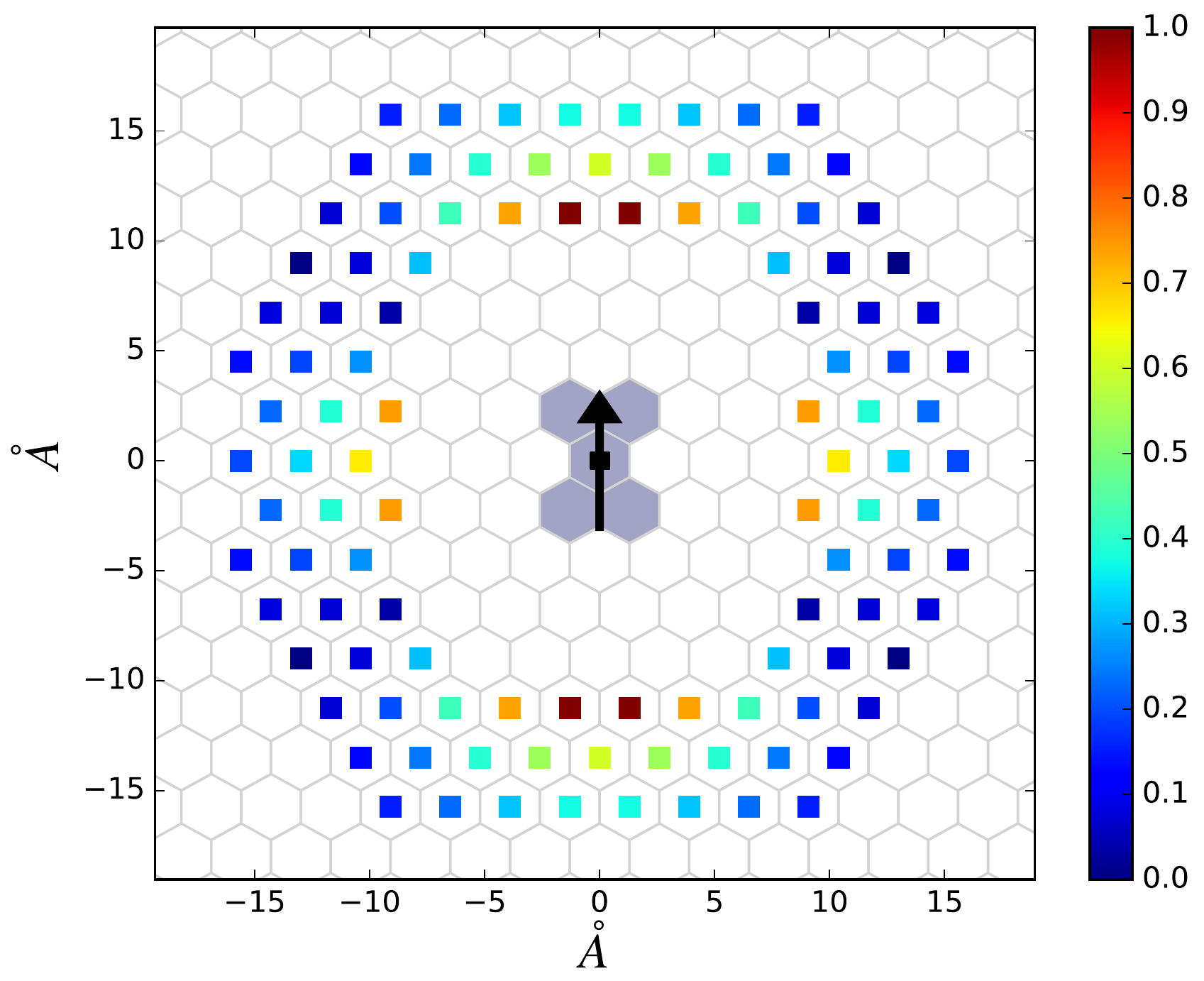}\label{fig:Jpp90_map}}
  \subfloat[Couplings in series 1 for three orientations]{\includegraphics[width=0.48\textwidth]{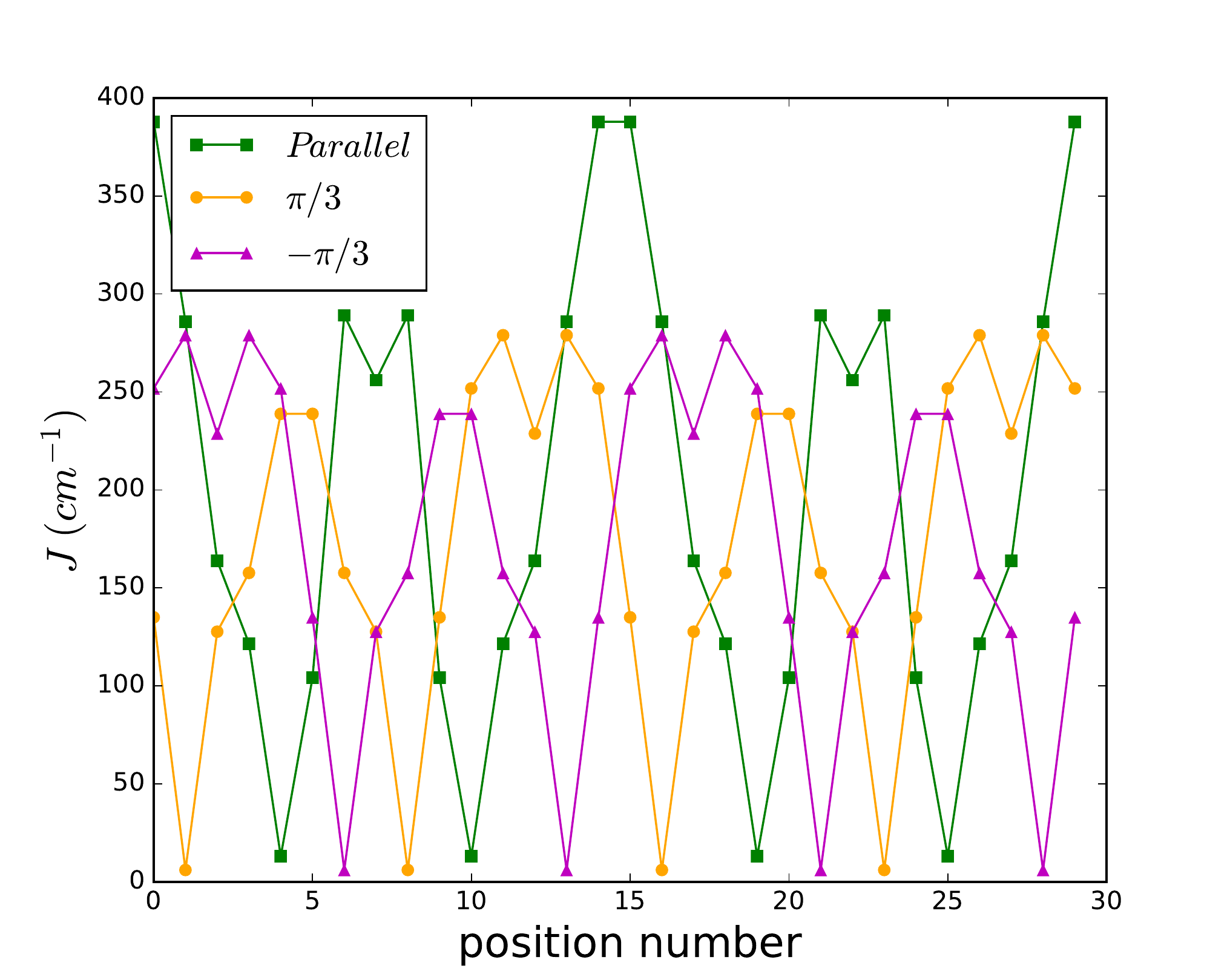}\label{fig:Jpp_l1_orientation}}
  \caption{(a) Map of resonance coupling between two parallel perylene moleculoids on the FG lattice. First moleculoid is at the center. Points show the positions of the center of the second moleculoid. Magnitude of the coupling $J$ is scaled to the largest value ($385.7\ \text{cm}^{-1}$) among all the couplings. (b) Magnitude of coupling in series 1 (closest series of cells to the center) as a function of orientation of the second moleculoid. The horizontal axis (position number) counts the cells counterclockwise and starting from the top. Three plots correspond to three relative orientations with respect to the original dipole moment with angles as written. Note the symmetry between $-\pi/3$ and $\pi/3$ orientations.}
\end{figure}

We compared excitation properties of perylene and anthanthrene moleculoids in FG with the vacuum properties of corresponding perylene and anthanthrene molecules. In vacuum, we use the geometry of an isolated defect optimized within FG sheet with only hydrogens added after the optimization to replace missing bonds between defect and the FG. The positions of hydrogens were optimized keeping all carbons frozen in FG geometry. Results from excited state calculation of both molecules (perylene and anthanthrene) are summarized in SI. The difference in values of transition energies and dipoles calculated for the molecular geometry optimized in FG  and the vacuum optimized geometry are on the order of only few per cent. Symmetry of the transition densities and orientations of transition dipoles are the same for all studied structures of isolated individual molecules and their respective FG defects. Transition dipoles for the isolated molecules are always smaller than the ones corresponding to moleculoids in the FG sheet. Magnitude of this enhancement of dipole moment for defect on FG depends on the size and shape of the defect (see SI).

\begin{figure}
\centering
\begin{minipage}{.47\textwidth}
  \centering
  \includegraphics[width=.9\linewidth]{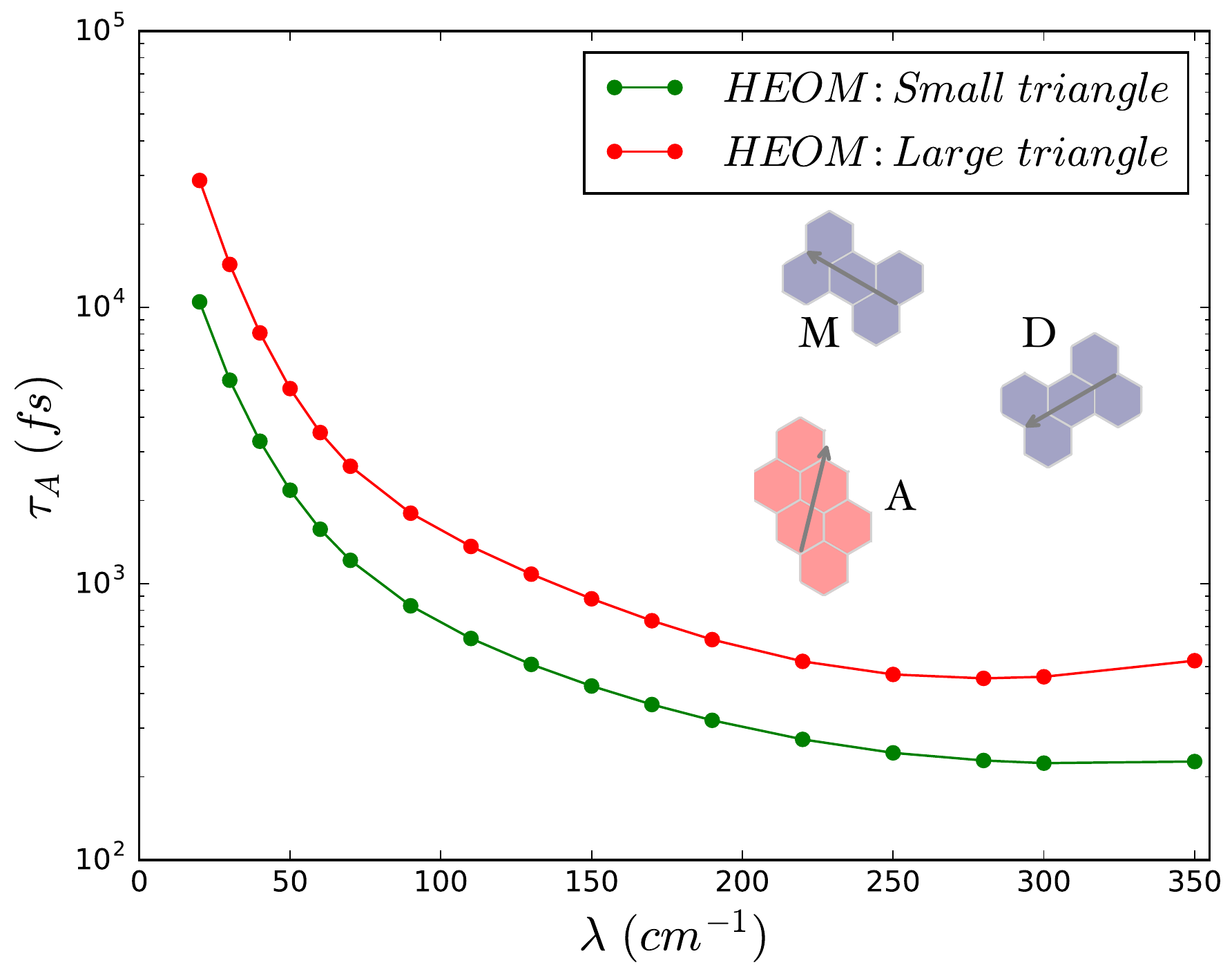}
  \captionof{figure}{Population transfer time for triangular arrangement of moleculoids. Population of the acceptor happens faster in the smaller triangle for a broad range of reorganization energies $\lambda$. Transfer time starts rising at the largest values of reorganization energy.}
  \label{fig:SB_taus_tri}
\end{minipage}%
\hfill
\begin{minipage}{.47\textwidth}
  \centering
  \includegraphics[width=.9\linewidth]{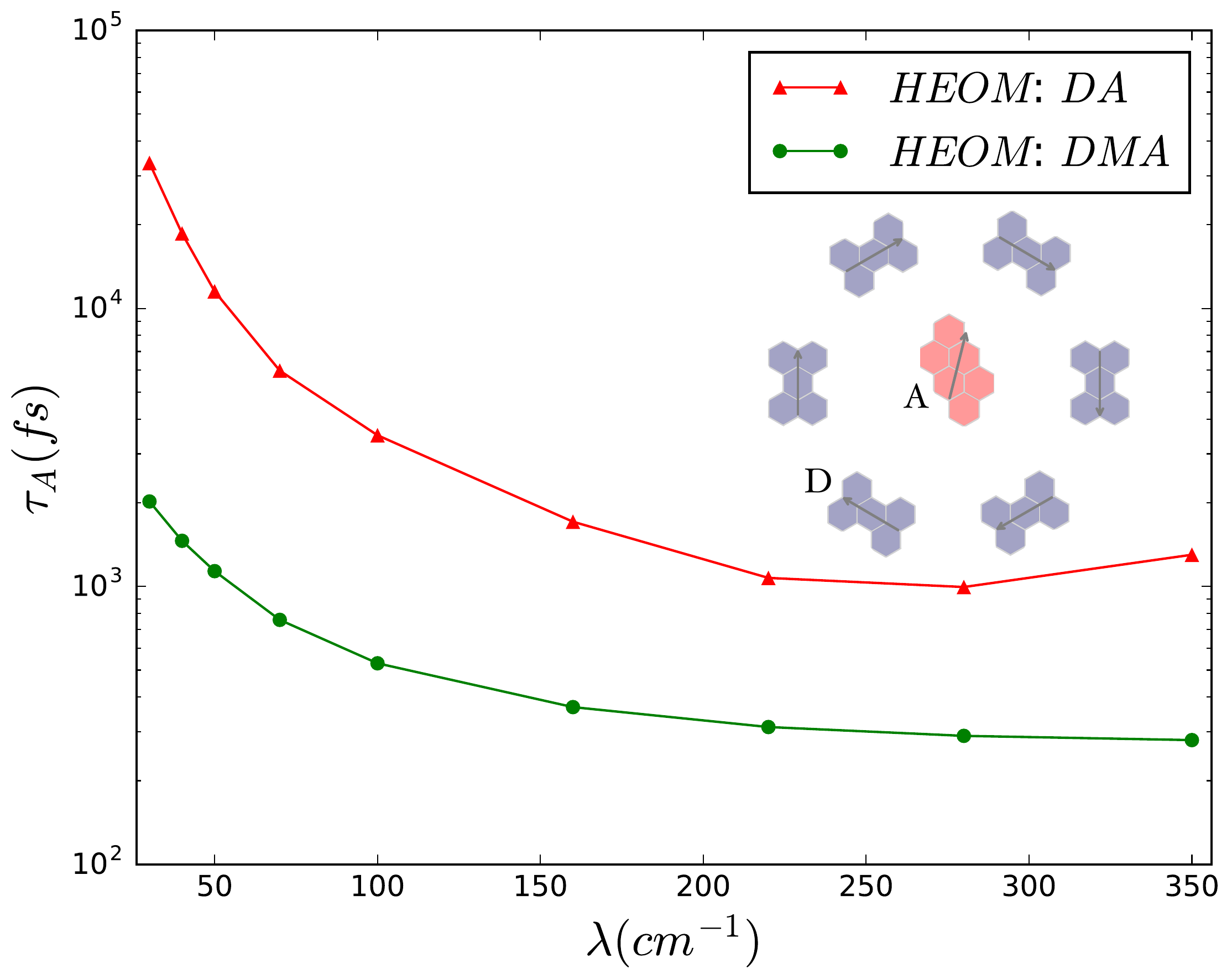}
  \captionof{figure}{Transfer time for both DMA and DA aggregates in hexagonal shape antenna. As is expected, energy transfer is considerably faster when the Middle conveyor (other five perylene moleculoids) is present.}
  \label{fig:hex151}
\end{minipage}
\end{figure}

For our selected moleculoids, we verify that to describe their mutual interaction it is sufficient to consider scaled dipole-dipole approximation. Hence,
\begin{equation}
\label{eq:dipole-dipole}
J_{ij} = \frac{\kappa}{4\pi\epsilon_0} \frac{\vec d_i \cdot \vec d_j - 3(\vec d_i \cdot \hat{r})(\vec d_j \cdot \hat{r})}{r^3},
\end{equation}
where $\vec d_i$ is the transition dipole moment between the ground and first excited states of moleculoid $i$, and $\hat{r}$ is the unit vector along $\vec{r}_{ij}$. The scaling factor $\kappa$ represents the effect of the FG environment on the interaction. It can also be understood as the inverse of effective relative permittivity $\epsilon_r^{eff}$ of the FG sheet.
To confirm the validity of this approximation, we compare couplings $J_{ij}$ calculated using Eq. \eqref{eq:dipole-dipole} with the resonance couplings calculated by QC methods for few distances.
According to the Frenkel exciton model, electronic coupling between two chromophores in a homodimer can be calculated as a half of the excited state splitting. We used this method for calculation of electronic coupling for two perylene moleculoids in FG surface with different mutual orientations and distances. This approach was chosen because there is no other straightforward way to include unknown FG effects into the calculation of interaction energy by standard methods such as transition density cube method \cite{krueger1998calculation} or Poisson-TrEsp method \cite{madjet2006intermolecular}. 
Structures with the defects were created from finite FG sheets in ideal periodic geometry (C-C distance 1.594 \AA \cite{zbovril2010graphene}) leaving always at least three rows of fluorinated graphene between the moleculoid and the edges of the FG sheet, to minimize the effects of its finite dimension.
For a few discrete perylene dimer configurations with parallel and (almost) serial dipole moments which differ in distances between their centers ($r$), we examine how couplings obtained from QC calculations change with $r$. Figure \ref{fig:rel_perm_a} clearly shows that QC resonance couplings linearly change with $r^{-3}$ which is a confirmation of validity of the dipole-dipole approximation for the moleculoids selected in our study. 

Assuming this approximation and couplings calculated by QC, Eq. \eqref{eq:dipole-dipole} determines the scaling factor $\kappa$ which characterizes the influence of the FG sheet on the coupling between moleculoids. For each possible mutual orientation of moleculoids we calculate resonance coupling $J_{ij}$ using transition dipoles calculated by QC for isolated moleculoids in vacuum (in FG geometry) as described earlier. Figure \ref{fig:rel_perm_b} shows relatively small differences in values of $\epsilon_r^{eff}=1/\kappa$. We estimate the average scaling factor for the FG sheet to be $\kappa = 1.43$ ($\epsilon_r^{eff} = 0.7$). Note that the decrease of $\epsilon_r^{eff}$ with the distance does not affect the resonance coupling as much as does the increase of distance $r^3$. Hence, the most relevant values for $\epsilon_r^{eff}$ to be averaged are the ones corresponding to shorter distances. Interestingly, the value of $\epsilon_r^{eff} < 1$  means that the interaction between moleculoids in FG is effectively enhanced with respect to the interaction between corresponding molecules in vacuum. This effect, partially a consequence of FG two-dimensionality, can be quantitatively explained, which we will do elsewhere.   

Exciton delocalization is crucial for achieving fast energy transfer in space. Hence, one of the most important questions for the utility of moleculoids in FG for light harvesting is whether the resonance couplings between them are strong enough to achieve exciton delocalization for moleculoid distances that satisfy zero deferential overlap condition. We study how the resonance coupling between two moleculoids varies with the position and relative orientation of the dipole moments within small distances. On the FG lattice, the angle between dipole moments of two perylene moleculoids takes values of $0, \frac{\pi}{3}, \frac{2\pi}{3}$ and $\pi$. For a given (upward pointing) perylene moleculoid, we show the three closest \textit{series} of locations for the center of a parallel perylene moleculoid in Figure \ref{fig:Jpp90_map}. Colors show the magnitude of resonance coupling $J$ which takes values $12.9 - 385.7\ \text{cm}^{-1}$ in series 1 (closest), $29.0 - 232.3\ \text{cm}^{-1}$ in series 2 (middle), and $2.0 - 143.3\ \text{cm}^{-1}$ in series 3 (furthest). The coupling as a function of the orientation of the second perylene dipole moment is plotted in Figure \ref{fig:Jpp_l1_orientation} for the first series of positions. It is clear from this plot that relative orientation of the two dipole moments is a crucial parameter for the value of resonance coupling. The values of couplings are in the range similar to the ones found in photosynthetic systems (see e.g. Fenna-Mathews-Olson complex \cite{van2000photosynthetic, Cho2005}, bacterial reaction center \cite{Jordanides2001}, LH2/3 complex \cite{Zigmantas2006} and many other systems in the literature). We construct our antenna from systems of equal or similar transition energy. Despite possible dynamic localization, the coupling values just noted are large enough for the delocalization of excited states to occur for reorganization energies of up to several hundreds of cm$^{-1}$. 

\begin{figure}[tbp]
\centering
\hspace*{-27em}
\subfloat[Square antenna]{\includegraphics[width=0.33\textwidth]{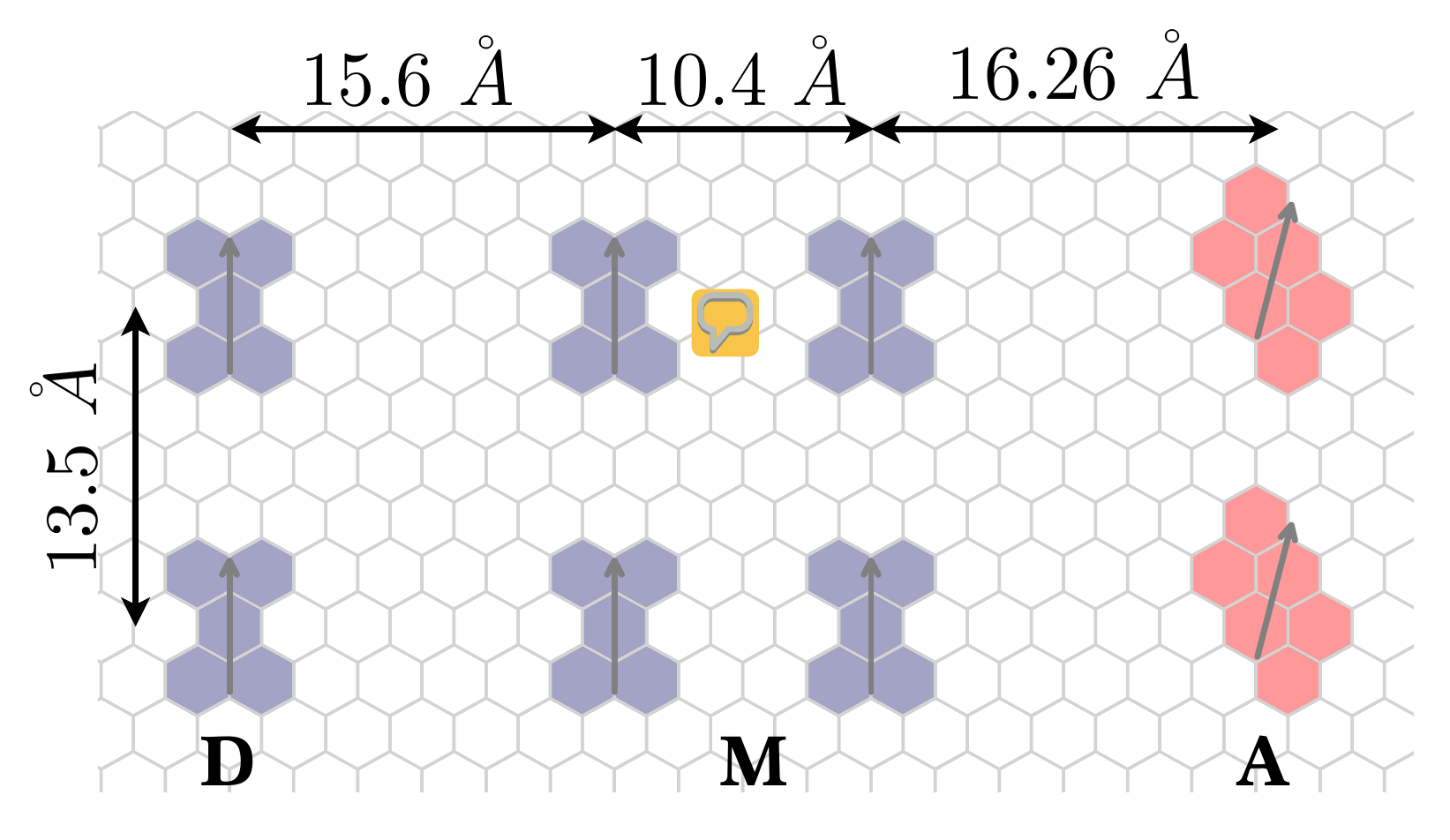}\label{fig:SquareDMA}}\\
\vspace*{-10.7em} \hspace{-1.8em}
\subfloat[Loop antenna]{\includegraphics[width=0.33\textwidth]{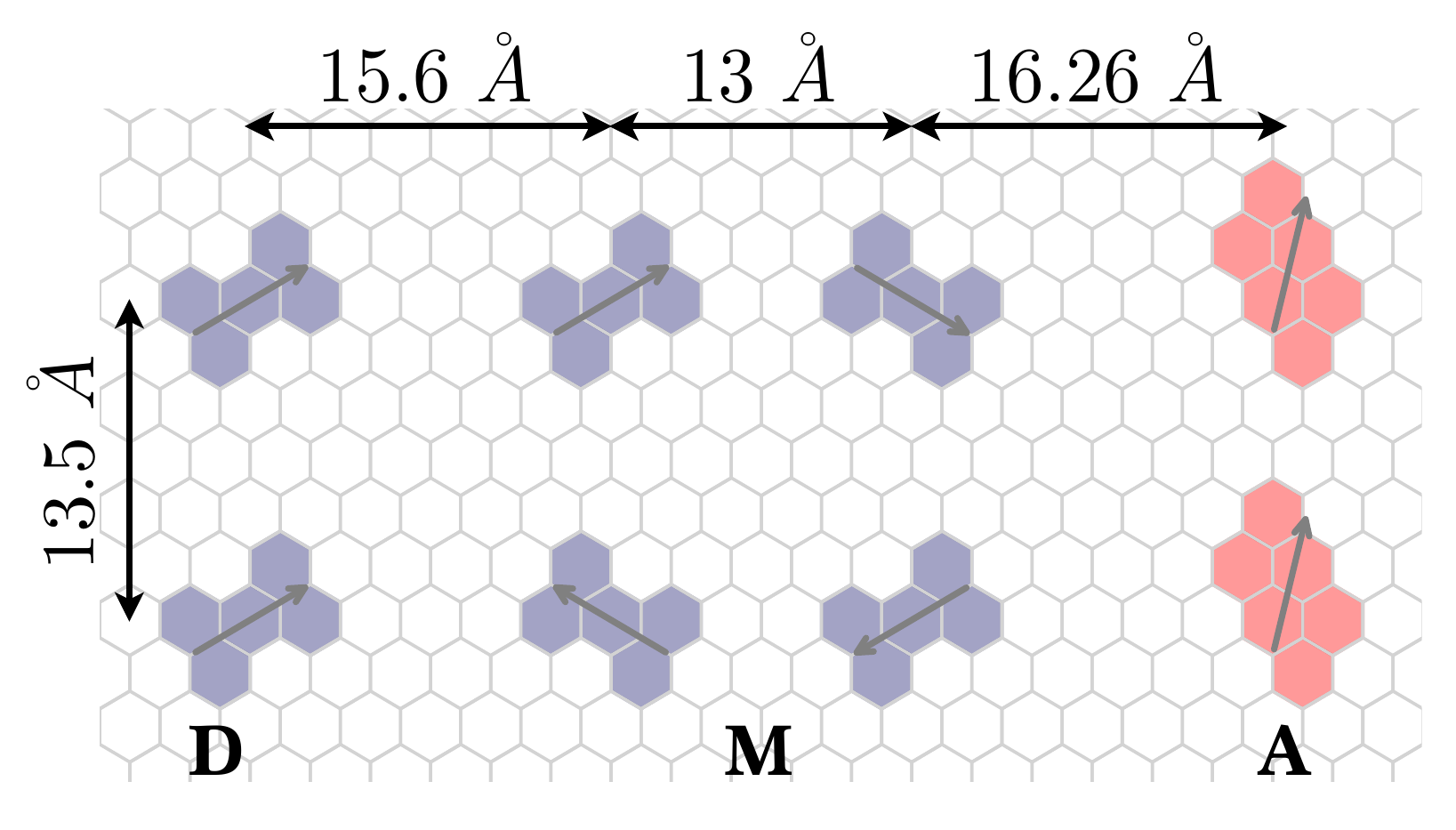}\label{fig:LoopDMA}}
\vspace*{-1em}
\subfloat[]{\includegraphics[width=0.5\linewidth]{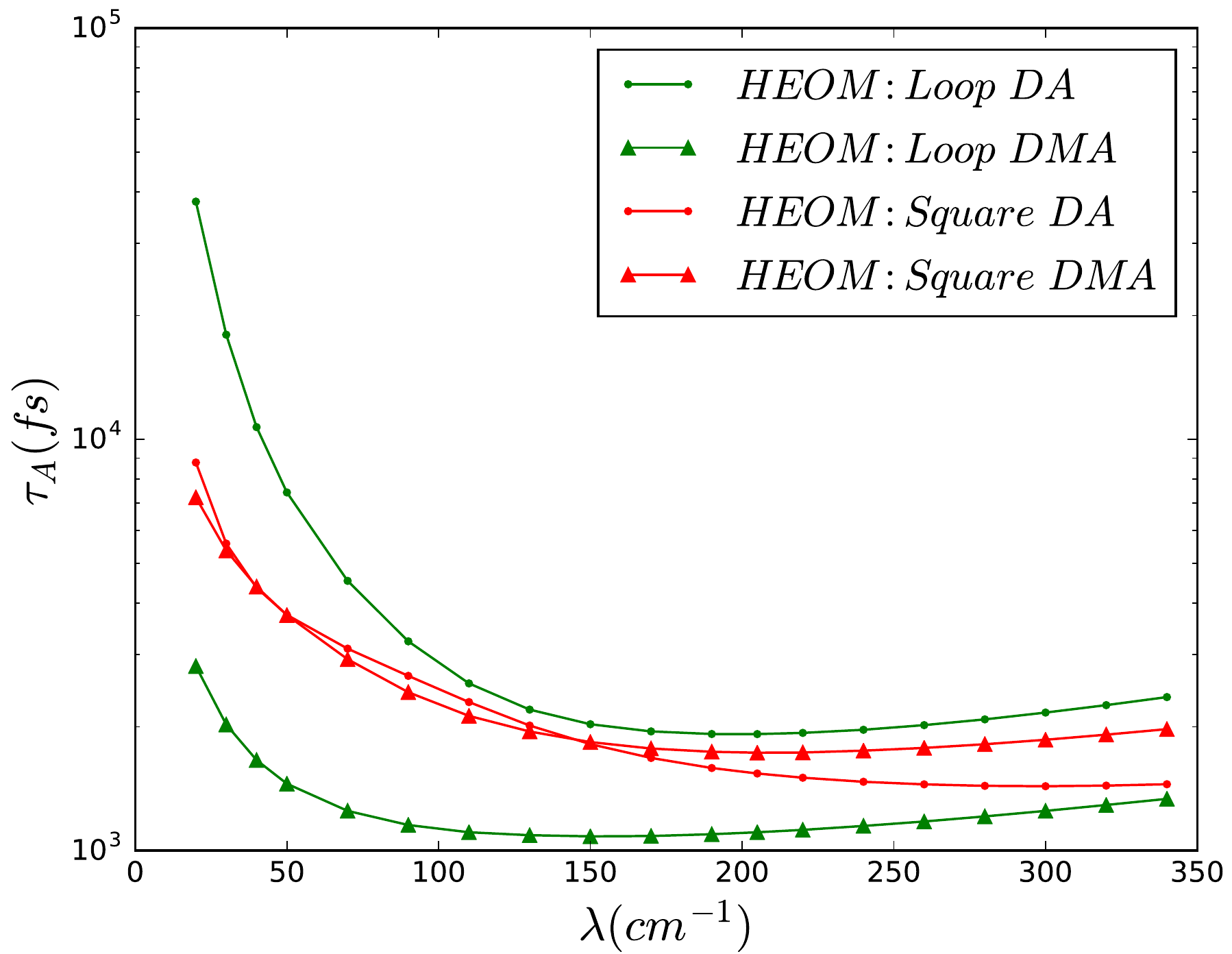}\label{fig:tau_rg_LS}}
\caption{Square (a) and Loop (b) antennae. (c) Acceptor transfer time $\tau_{A}$ in the 8-site aggregate of moleculoids with two different configurations for the donor and the middle conveyor. Faster energy transfer through the DMA than DA arrangement is found in Loop antenna for all reorganization energies $\lambda$, and for some values of $\lambda$ in Square antenna, compared to the corresponding DA arrangements.}
\label{fig:LoopSquare}
\end{figure}

\subsection*{Energy Transfer in Model Antennae}
In order to demonstrate the central ideas behind our proposed artificial light-harvesting systems, we construct several examples of artificial antennae for energy transport from a donor molecule to an energetically lower acceptor molecule. We choose the donor (D) moleculoid to be of the perylene type, and the acceptor (A) moleculoid of the anthanthrene type, with excitation energies $ E_P = 22354 \ \text{cm}^{-1}$ and $ E_A = 21736 \ \text{cm}^{-1}$, respectively, obtained from QC calculations. The energy gap between D and A ensures thermodynamically that preferential donor to acceptor energy transfer occurs, because a larger than $k_{B}T$ energy gap has to be crossed (for temperature $T$ around $300$ K). The transition dipole moments calculated for these moleculoids are 6.30 $D$ and 6.55 $D$, respectively (see SI). As in natural photosynthetic systems, donor and acceptor sites my be separated by considerable spatial and energy gaps. In order to aid the excitation in crossing these gaps, we add a middle exciton-conveyor aggregate (M) between the D and A molecules (Figure \ref{fig:DMA1}). Further on in this section, we study various geometries for the DMA assembly as well as different allowed orientations for each moleculoids on the lattice. We present some examples confirming the idea of energy funneling among our selected types of moleculoids. Moreover, the importance of the middle conveyor and its role in excitation transfer is demonstrated by comparing DMA aggregates to the corresponding ones only consisting of D and A. 
As a measure of transfer efficiency we calculate characteristic time $\tau_A$ during which the acceptor gets populated, and compare it for the cases of DMA and DA complexes (see Methods section). 

The crucial ingredient of efficient energy transfer in natural photosynthetic systems is the presence of environment which efficiently accepts excess electronic energy when excitation travels down the energy funnel. The action of this environment is embodied in energy transfer theories which describe limiting cases of weak and strong system-environment coupling of energy transfer, namely the Redfield (weak regime) and F\"orster (strong regime) theories \cite{van2000photosynthetic,valkunas2013molecular}. In both of these theories the rates of energy transfer directly depend on the presence of bath states which complement electronic energy to achieve resonance. The resonance condition for energy transfer reflects the energy conservation in the total antenna-plus-environment system.  In line with the most common description of photosynthetic environment, we assume linear couplings between electronic transitions and an infinite bath of (mostly) nuclear degrees of freedom. We assume that not only the FG sheet, but also its immediate environment, such as solvent or surface on which it is deposited, contribute to the bath degrees of freedom. Although very little can be said {\it a priori} about the particular strength of coupling between moleculoid transitions and their environment in FG, one can take the liberty of exploring certain broad range of parameters to see under which conditions fast energy transfer can occur. 

To model the environmental influence, we assume that the environment has an overdamped Brownian oscillator spectral density $J(\omega) = \frac{2\lambda\gamma\omega}{\omega^2 + \gamma^2}$,  where $\lambda$ and $\gamma$ are reorganization energy and inverse correlation time, respectively. We take $\gamma = 1/60$ fs$^{-1}$ in all calculations, and vary the reorganization energy to obtain the range of achievable transfer times. The transfer time is our best proxy to efficiency of energy transfer. Transfer efficiency can be defined as a ratio of the number of excitations that reach the acceptor to the total number of excitations. This ratio can be expressed through the inverse excitation life-time, which is in nanoseconds for natural chlorophyll based light-harvesting systems, and the time that excitation spends in the antenna before reaching its final destination. In the case of FG moleculoids, excitation life-time is unknown. However, the radiative part of this life-time is determined by the transition dipole strength, and it is therefore expected to be of similar order of magnitude as in chlorophyll based systems. Intersystem crossing to triplet states is expected to have the same nanosecond time scale as for corresponding molecules. The remaining mechanisms of non-radiative depopulation of excited states, such as conical intersections, often involve large changes of molecular structure. These could also be excluded for moleculoids whose nuclear configuration is to a large extent determined by the FG sheet. Correspondingly, one can assume moleculoid excited state life-times in the rage of nanoseconds, similar to the values known for the corresponding molecules in gas phases \cite{Ware1966}. This means that excitation transfer times in the range of tens of picoseconds will lead to transfer efficiency over $90\%$.

Through the examples that follow, we show how the exciton energy transfer may change for different structures of light harvesting antenna. We confine our example antennae to small clusters -- up to 8 moleculoids -- for which we calculate exact population dynamics by hierarchical equations of motion (HEOM)\cite{tanimura1989time,ishizaki2009unified,tanimura2012reduced}. Initial state for the propagation is a unity population on one of the sites (the donor site) of the system. We investigate a broad range of values of reorganization energies $\lambda$ from $20$ to $350 \ \text{cm}^{-1}$. All examined antennae show that the acceptor sites get populated exponentially with time. We read off the characteristic time of the acceptor ($\tau_A$) from a single exponential fit, to estimate the overall speed of excitation energy transfer in the model antennae.

As the simplest example, we consider \textit{triangle} configuration of one anthanthrene moleculoid (A) and two perylene moleculoids (D and M) with similar mutual distances so that the couplings in A-D, M-A and D-M have similar values. We study a `small' and a `large' \textit{triangle} to realize the effect of the inter-monomer distances on the dynamics of exciton transfer. D-M, M-A and A-D distances in the two cases are $13.5, 12.53, 14.75$ \AA\ (small) and  $18.00, 16.99, 19.22$ \AA\ (large), respectively. The values of resonance coupling are $49, 55$ and $48 \ \text{cm}^{-1}$ in the large triangle, and $116, 120$ and $99 \ \text{cm}^{-1}$ in the small one. For all studied values of the reorganization energy, the smaller antenna shows faster energy transport to the acceptor site (Figure \ref{fig:SB_taus_tri}). The transfer time decreases with increasing reorganization energy. The longest transfer times are in tens of picoseconds. Even for the slower antenna they drop under hundreds of femtoseconds with increasing $\lambda$.  

In the next antenna, perylene moleculoids are placed on the FG lattice in a hexagonal configuration. The distance between them is taken $5b=13$ \AA\, and the initial population is carried by one of these moleculoids as shown in Figure \ref{fig:hex151}. Acceptor is placed in the middle of the hexagon. The resonance couplings between moleculoids take values between $16$ and $228\ \text{cm}^{-1}$. HEOM computations for population transfer are performed for reorganization energies in the range of $30-350\ \text{cm}^{-1}$, clearly showing that the transfer is slower when the system consists of only donor and acceptor: the larger values of $\tau_A$ for DA in Figure \ref{fig:hex151} for the whole range of $\lambda$. This case demonstrates a general fact that combining $N$ equivalent donors in a favorable configuration, while increasing absorption cross-section $N$-fold, may simultaneously increase energy transfer rate from each of these donors to the acceptor.  

Figures \ref{fig:SquareDMA} and \ref{fig:LoopDMA} present two larger systems. Based on the orientation of the moleculoids in their middle conveyor we denote them as Loop (L) and Square (S). For each antenna, we calculate the population transfer time to the acceptor through M, and compare it to the transfer times in the corresponding DA configuration, while the acceptor is  placed at half the original distance from the donor, i.e. in the middle of space originally taken by M. Figure \ref{fig:tau_rg_LS} shows that in each aggregate and particularly for the smaller values of the reorganization energy, population transfer is considerably faster when M is present, despite of the larger donor-acceptor distance. Moreover, one can compare the energy transfer rate between the two antennae in order to assess the role of geometry: orientation of the transition dipole moments (locations are almost the same). Loop antenna shows faster energy transfer over the whole values of reorganization energy compared to the Square antenna. The differences are considerable, meaning that the orientation of dipole moments can hugely influence the dynamics. Therefore, the moleculoid orientation seems to be a major factor for the optimized antenna. Another interesting feature of the transfer time dependence on reorganization energy is that as $\lambda$ increases to the largest values, the antennae show slightly longer transfer times. This is a consequence of a competition between resonance coupling and coupling to the environment which results in dynamic localization. With fixed resonance couplings, weak system-environment coupling theory (Redfield theory) shows a linear increase of energy transfer rate with reorganization energy \cite{van2000photosynthetic,may2008charge}. Breakdown of linearity signals that system-environment coupling cannot be considered weak any more, and that its renormalization effect on the system Hamiltonian cannot be ignored. 

The values of transfer time $\tau_A$ in all the studied small model antennae are between hundreds of femtoseconds and tens of picoseconds. These times are in a region favorable to efficient light-harvesting. Using tight arrangement of larger number of moleculoids could result in even faster energy transfer, and quantum efficiency as large as the ones found in the best performing natural light-harvesting systems. Exact treatment of model antennae of arbitrary size interacting with a general environment is not currently possible. HEOM method cannot cheaply handle arbitrary spectral densities, and for overdamped Brownian spectral density it can only handle a limited number of pigments. However, to study larger arrangements of moleculoids in a reliable approximative fashion, one can use the standard perturbative methods. Especially in the case that the system can be split into groups of moleculoids with strong resonance coupling inside the group and weak coupling between the groups, a combination of Redfield theory and the so-called multi-chromophoric F\"orster theory \cite{Sumi1999, Jang2004, Scholes2001} yields rather reliable results. The quality of these methods for particular cases can be assessed by comparison with HEOM. For instance, for an antenna with similar or identical chromophores, such as our middle conveyor, Redfield theory gives reliable results for populations of sites even in the regime where reorganization energy is comparable with resonance coupling, i.e. outside its strict regime of validity (see SI).  
 
\section*{Discussion}
In this work, we present the idea of utilizing graphene-like defects or impurities in fluorographene for light energy harvesting and spatial and energy domain excitation energy transfer. We advocate a molecular approach to this interesting material, as the defects we study exhibit properties which are remarkably molecule-like, hence the term {\it moleculoid}. We study sections of a single fluorograhene sheet with two small types of moleculoids. We test the utility of quantum chemical methods for calculation of excited state energies, transition dipole moments, optimal geometry and energetic stability. We verify that scaled dipole-dipole formula can be used to calculate one of the crucial parameters of energy transfer theory, the resonance coupling between pairs of moleculoids. We applied Frenkel exciton model to calculate the relevant energy level structure of small aggregates of moleculoids which represent our model light-harvesting antennae. One of the conditions of application of Frenkel exciton model is the lack of differential overlap of molecular orbitals between neighbouring molecules. In the context of efficient excitation energy transfer, this condition can be regarded as one of the design principles of efficient light-harvesting antenna. We translate this condition to the case of fluorographene and verify by quantum chemistry calculations that small moleculoids satisfy this condition rather well even for the minimum distance in which they can be naturally considered separate. Resonance couplings for perylene and anthanthrene moleculoids constrained in their positions and orientations by fluorographene lattice are found large enough to provide exciton delocalization, one of the key conditions of fast energy transfer. We construct several model antennae and show that for reasonable range of system-environment coupling strengths (measured by reorganization energy) we obtain energy transfer rates which would result in high quantum yields in presence of expected losses. We thus expect that construction of efficient artificial light-harvesting antennae using fluorographene moleculoids is feasible.

A broader concept of moleculoid antennae includes the possibility to tune antenna properties by combination of different types of moleculoids. Larger moleculoids are expected to absorb light on longer wavelengths and similar designs could thus be translated to different wavelength regions for simultaneous light-harvesting of different sections of solar spectrum. Also, it did not escape our attention that natural photosynthesis is a process occurring on membranes. Moleculoid systems are naturally equipped with a membrane, the fluorographene itself. The system of antennae on a fluorographene sheet could thus be naturally made a part of a true photo-synthetic machinery in which, like in natural photosynthesis, charges are pumped across a thin membrane to be utilized in chemical synthesis. Last but not least, the concept of $\pi$-conjugated isles in 2D materials is not restricted to fluorographene. Other 2D materials, e.g. other modifications of graphene, may provide similar opportunities. Obvious first guess would be graphane and materials using other halogen atoms -- these seem however not to be stable without further modifications. The idea is also not restricted to a single sheet. The prospect of building membrane with several interacting 2D sheets and molecuoids interacting in three dimensions is also worth investigating.  

Recently, we have identified naturally occurring candidate moleculoids in the band of nearly pure fluorographene by single molecule spectroscopy \cite{malyInPrep}. Single molecule measurements confirm natural occurrence of moleculoids of somewhat larger types than the ones studied in this work. Assignment and detailed properties of other types of moleculoids will be studied elsewhere.  

\section*{Methods}

\subsection*{Quantum Chemistry}
Quantum chemistry calculations were performed using Gaussian 09 \cite{gaussian}. Structure of all studied systems was optimized using DFT approach with BLYP functional and LANL2DZ valence double zeta basis set. Excited state properties were obtained by TDDFT approach with $\omega$B97XD long range corrected hybrid functional and LANL2DZ basis set. This combination of methods for geometry optimization and excited state calculation was chosen by comparison of transition energies obtained by combination of different functionals and basis sets with experimental results on testing set of 15 aromatic hydrocarbons (see SI). Our criterion was to obtain the same accuracy, i.e. the same systematic shift of the excited states from experimental values, for different hydrocarbon sizes and shapes, rather than the best agreement with the experiment on some smaller subset. This way we can assume that our approach can reliably describe wide range of hydrocarbon molecules and molecule-like defects on FG.

\subsection*{Open Quantum Systems Theory}

Site population dynamics of our model antennae were calculated using GPU-HEOM tool available on-line \cite{GPUHEOM_nanohub}. We limited the calculations to the truncation level 4 of Kubo-Tannimura hierarchy, which also shows good agreement with truncation level 5. For all calculations we used time step of $1 \ \text{fs}$. Sites are coupled to independent vibronic baths which are characterized by the spectral density $J(\omega) = \frac{2\lambda\gamma\omega}{\omega^2 + \gamma^2}$ as was explained above. We vary the reorganization energy $\lambda$ over a broad range of values from weak to strong couplings and obtain the population transfer for each value. 

\subsection*{Estimation of Transfer Time}
The transfer time $\tau_A$ from the antenna to the acceptor is determined by fitting the population of the acceptor in a given antenna by the formula $P_{A}(t) \approx a(1 - e^{-t/\tau_A})$.
Values of $a$ and $\tau_A$ are found using least-square fitting implemented in Scipy Python package \cite{scipy}. 


\section*{Acknowledgements}

This work was supported by the Neuron Fund for Support of Science, grant Neuron Impuls for Physics 2014. V.S. acknowledges financial support of Grant Agency of Charles University (GAUK) grant no. 1162216. Access to computing and storage facilities owned by parties and projects contributing to the National Grid Infrastructure MetaCentrum provided under the programme ``Projects of Large Research, Development, and Innovations Infrastructures" (CESNET LM2015042), is greatly appreciated.

\end{document}